\documentclass[aoas,MSNbibl,nameyear,dvips]{arximspdf}
\usepackage{graphicx}

\doi{10.1214/14-AOAS786} 
\volume{8}
\issue{4}
\pubyear{2014}
\firstpage{2509}
\lastpage{2537}
\docsubty{FLA}

\makeatletter
\newcommand{\est}{\operatorname{est}}
\newcommand{\Var}{\operatorname{Var}}
\newcommand{\bbm}[1]{\mathbf{#1}}
\newcommand{\bmm}[1]{\bolds{#1}}
\makeatother

\begin{document}
\begin{frontmatter}

\title{Reduced-rank spatio-temporal modeling of air pollution
concentrations in the Multi-Ethnic Study of Atherosclerosis and Air
Pollution\thanksref{T1}}
\runtitle{Reduced-rank spatio-temporal modeling of air pollution}

\begin{aug}
\author[A]{\fnms{Casey}~\snm{Olives}\corref{}\thanksref{m1}\ead[label=e1]{colives@uw.edu}},
\author[B]{\fnms{Lianne}~\snm{Sheppard}\thanksref{m1}\ead[label=e2]{sheppard@uw.edu}},
\author[C]{\fnms{Johan}~\snm{Lindstr\"{o}m}\thanksref{m2}\ead[label=e3]{johanl@maths.lth.se}},
\author[D]{\fnms{Paul~D.}~\snm{Sampson}\thanksref{m1}\ead[label=e4]{pds@stat.washington.edu}},
\author[A]{\fnms{Joel D.}~\snm{Kaufman}\thanksref{m1}\ead[label=e5]{joelk@uw.edu}}
\and
\author[B]{\fnms{Adam A.}~\snm{Szpiro}\thanksref{m1}\ead[label=e6]{aszpiro@uw.edu}}
\runauthor{C. Olives et al.}
\affiliation{University of Washington\thanksmark{m1} and Lund University\thanksmark{m2}}
\address[A]{C. Olives\\
J. D. Kaufman\\
Department of Environmental\\
\quad and Occupational Health Sciences\\
University of Washington\\
4225 Roosevelt Way NE\\
Seattle, Washington 98105\\
USA\\
\printead{e1}\\
\phantom{E-mail: }\printead*{e5}}
\address[B]{L. Sheppard\\
A. A. Szpiro\\
Department of Biostatistics\\
University of Washington\\
Box 357232\\
Health Sciences Building Room F600\\
1705 NE Pacific\\
Seattle, Washington 98195-7232\\
USA\\
\printead{e2}\\
\phantom{E-mail: }\printead*{e6}}
\address[C]{J. Lindstr\"{o}m\\
Mathematical Statistics\\
Center for Mathematical Sciences\\
Lund University\\
Box 118, SE-221 00 Lund\\
Sweden\\
\printead{e3}}
\address[D]{P. D. Sampson\\
Department of Statistics\\
University of Washington\\
Seattle, Washington 98195-4322\hspace*{18pt}\\
\printead{e4}}
\end{aug}
\thankstext{T1}{Supported in part by Grants T32ES015459 and K24ES013195
from the National Institute of Environmental Health Sciences of the
National Institutes of Health (NIH). Additional support was provided by
the U.S. Environmental Protection Agency (EPA), Assistance Agreement
\mbox{RD-83479601-0} (Clean Air Research Centers) and CR-834077101-0. This
publication was developed under a STAR research assistance agreement,
No. RD831697, awarded by the U.S Environmental Protection Agency.}

\received{\smonth{10} \syear{2013}}
\revised{\smonth{8} \syear{2014}}

\begin{abstract}
There is growing evidence in the epidemiologic literature of the
relationship between air pollution and adverse health outcomes.
Prediction of individual air pollution exposure in the Environmental
Protection Agency (EPA) funded Multi-Ethnic Study of Atheroscelerosis
and Air Pollution (MESA Air) study relies on a flexible spatio-temporal
prediction model that integrates land-use regression with kriging to
account for spatial dependence in pollutant concentrations. Temporal
variability is captured using temporal trends estimated via modified
singular value decomposition and temporally varying spatial residuals.
This model utilizes monitoring data from existing regulatory networks
and supplementary MESA Air monitoring data to predict concentrations
for individual cohort members.

In general, spatio-temporal models are limited in their efficacy for
large data sets due to computational intractability. We develop
reduced-rank versions of the MESA Air spatio-temporal model. To do so,
we apply low-rank kriging to account for spatial variation in the mean
process and discuss the limitations of this approach. As an
alternative, we represent spatial variation using thin plate regression
splines. We compare the performance of the outlined models using EPA
and MESA Air monitoring data for predicting concentrations of oxides of
nitrogen (NO$_x$)---a pollutant of primary interest in MESA Air---in
the Los Angeles metropolitan area via cross-validated $R^2$.

Our findings suggest that use of reduced-rank models can improve
computational efficiency in certain cases. Low-rank kriging and thin
plate regression splines were competitive across the formulations
considered, although TPRS appeared to be more robust in some\vspace*{-6pt} settings.
\end{abstract}

\begin{keyword}
\kwd{Spatiotemporal modeling}
\kwd{reduced-rank}
\kwd{air pollution}
\kwd{kriging}
\kwd{thin plate splines}
\end{keyword}
\end{frontmatter}

\section{Introduction}

There is growing evidence in the epidemiologic literature of the
relationship between air pollution and adverse health outcomes. Early
findings were based on somewhat crude regional, and possibly temporally
specific, assignment of exposures [\citet{Dockery1993uq,Pope2002vn,Samet2000kx}]. Yet, methods for assigning
individual exposure to cohort study participants have become much more
sophisticated. Recent studies have assigned individual exposure using
the value measured at the nearest monitoring location [\citet{Miller2007ly,Ritz2006zr}]; using ``land use regression'' estimates
based on spatially distributed or Geographic Information Systems (GIS)
based covariates [\citet{Brauer2003ys,Hoek20087561,Jerrett2005fk}];
and by interpolation with geostatistical methods such as kriging and
semi-parametric smoothing [\citet{Jerrett2005qf,Kunzli2005ve,Paciorek2009}].

Motivated by the Multi-Ethnic Study of Atheroscelerosis and Air
Pollution (MESA Air) study [\citet{Kaufman2012}], \citet{Szpiro2009},
\citet{Sampson2011} and \citet{Lindstrom2012} developed a flexible
spatio-temporal prediction model based on monitoring data from existing
regulatory networks as well as supplementary MESA Air monitoring data
to predict concentrations for individual MESA cohort members. This work
integrates land-use regression with kriging to account for spatial
dependence in pollutant concentrations. Temporal variability is
captured using temporal trends estimated via sparse singular value
decomposition and temporally varying spatial residuals [\citet{fuentes2006using,Sampson2011,Szpiro2009}].

In general, spatio-temporal models are limited in their efficacy for
large data sets due to computational intractability. For example, in
the purely spatial setting, computation typically is of the order
$\mathcal{O}(n^3)$, where $n$ is the number of spatial locations. The
computational effort for log-likelihood evaluation of the MESA Air
spatio-temporal model typically grows at least as fast, but slower than
$\mathcal{O}(N^3)$, where $N$ is the total number of spatio-temporal
observations [\citet{Lindstrom2012}]. Methods for reducing the
computational burden in spatio-temporal models are becoming more common
in the spatial statistics literature. Several authors have proposed
dynamic frameworks for modeling residual spatial and temporal
dependence, although these approaches continue to suffer from
computational intractability [\citet{Gelfand2005,Stroud2001}]. In the
large spatial data context, approximate likelihood and sampling-based
approaches have been proposed to reduce computational burden [\citet{Feuntes2007,Pace20091435-5930209}]. An alternative to approximate
methods involves reducing the spatial process to a $K$-dimensional
subspace ($K \ll n$) in order to increase computational efficiency [\citet
{Banerjee2008,Crainiceanu2008,Kammann2003zr,Nychka1998}, \citeauthor{Stein2007} (\citeyear{Stein2007,Stein20083})].
These so-called ``low-rank'' or ``reduced-rank'' approaches can reduce
computation to $\mathcal{O}(K^3)$.

\begin{figure}

\includegraphics{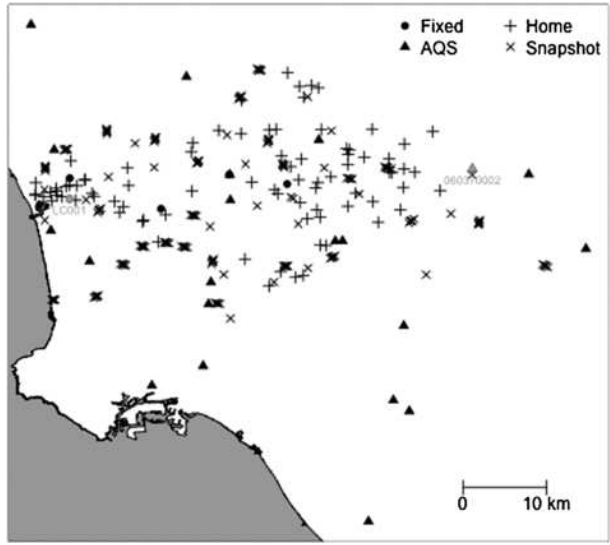}

\caption{Map of AQS and MESA Air monitoring locations in Los Angeles,
California. ``Home outdoor'' monitors have been jittered for participant
confidentiality.}
\label{figmapofdata}
\end{figure}

In the current work, we develop reduced-rank versions of the
spatio-temporal model outlined in \citet{Lindstrom2012,Szpiro2009}.
Specifically, we apply the approach proposed by \citet{Kammann2003zr}
to achieve low-rank kriging to account for spatial variation in the
mean process and spatially varying temporal trends. We discuss the
limitations of this approach and, as an alternative, represent spatial
variation using thin plate regression splines [\citet{Wood2003uq}]. We
compare the performance of the outlined models using Environmental
Protection Agency (EPA) and MESA Air monitoring data for predicting
oxides of nitrogen (NO$_x$) concentrations in the Los Angeles
metropolitan area.

\section{Description of data}

\subsection{Air Quality System (AQS)}

The national AQS network of regulatory monitors, managed by the EPA,
reports concentrations of a wide variety of air pollutant
concentrations on an ongoing basis, most typically hourly averages. For
this study, we include NO$_x$ measurements from 21 AQS monitors in the
Los Angeles area, one of six metropolitan areas where MESA Air cohort
members live. Monitor locations are shown in Figure~\ref{figmapofdata} (left). As MESA Air supplementary monitoring is done
at the 2-week average scale, we aggregate AQS monitoring data to 2-week
averages. Due to skew in the data, all 2-week averages are log transformed.

\subsection{MESA Air}

As part of the MESA Air project goals to provide high quality
individual exposure prediction, additional monitoring data were
collected in each of the study's six geographic regions, including Los
Angeles. The goal of the supplementary monitoring was to provide
geographically complementary data to the AQS monitoring data and to
systematically span the design space based on proximity to traffic.
Additionally, supplementary monitoring data included measurements
collected at a subset of cohort participant homes. The sampling
strategy is described in more detail by \citet{Cohen2009}.

\begin{figure}

\includegraphics{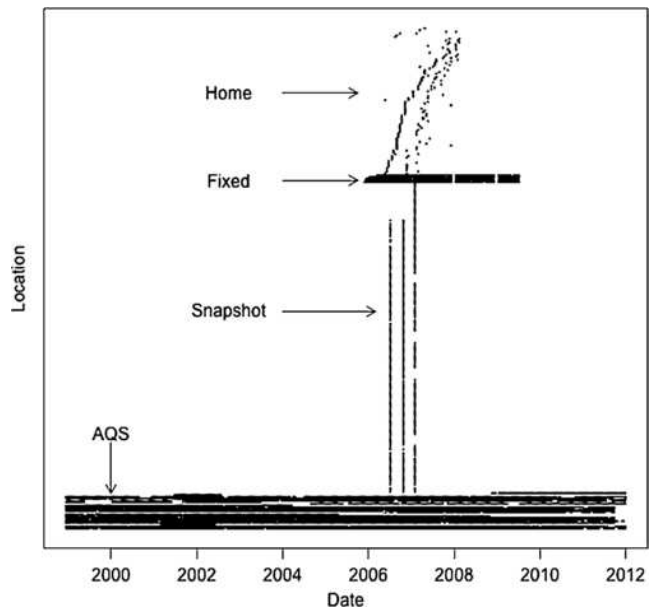}

\caption{Schematic of sampling schedule for AQS and MESA Air monitors
between 1999 and 2012. Each point represents a two-week sampling period.}
\label{figdateschematic}\vspace*{5pt}
\end{figure}

The MESA Air supplementary data is comprised of three classes of
monitors, which we refer to as ``fixed site,'' ``home outdoor'' and
``community snapshot.'' There are a total of five ``fixed sites''
included in this study in the Los Angeles area. These ``fixed-sites''
began measuring 2-week average concentrations in November of 2005, for
a total of 426 measurements by June 1, 2009. A total of 84\vadjust{\goodbreak} ``home
outdoor'' locations were included in this study. These sites were
sampled during 2-week periods starting in May of 2006 and ending in
February of 2008, for a total of 155 measurements. The sampling plan
calls on each home to be measured two times during different seasons.
Last, the ``community snapshot'' sub-campaign consists of 177 sites
measured in three rounds of spatially rich sampling during single
2-week periods from July 5, 2006 to January 1, 2007, for a total of 449
measurements. In each round of the ``community snapshot'' monitoring,
most monitors were clustered in groups of six, with three on each side
of a major roadway at distances of about 50, 100 and 300 meters, and
locations were chosen to span the domain of various land-use categories
and to cover a wide geographic region. All MESA Air monitoring
locations as of June 1, 2009 are displayed in Figure~\ref{figmapofdata}. Likewise, temporal coverage and sampling frequency
during the study period for each monitoring location and type is
depicted in Figure~\ref{figdateschematic}. Table~\ref{tabsummary}
provides summary statistics on the native and log-scales for both EPA
and MESA Air data.

\begin{table}[b]\vspace*{5pt}
\tablewidth=\textwidth
\caption{Summary of statistics of NO$_x$ monitoring data at EPA AQS and
MESA Air supplementary monitoring sites}
\label{tabsummary}
\begin{tabular*}{\tablewidth}{@{\extracolsep{\fill}}lcccc@{}}
\hline
& \multicolumn{2}{c}{$\mathbf{NO}_{\bolds{x}}$  \textbf{ppb}} & \multicolumn{2}{c@{}}{$\operatorname{\mathbf{log}}\textbf{(}\mathbf{NO}_{\bolds{x}}$ \textbf{ppb}\textbf{)}}\\[-4pt]
& \multicolumn{2}{c}{\hrulefill} &\multicolumn{2}{c@{}}{\hrulefill} \\
\textbf{Type} \textbf{of} \textbf{site} & \textbf{Mean} & \textbf{SD} & \textbf{Mean} & \textbf{SD}\\
\hline
AQS/fixed site &&&&\\
\quad 2-wk & 53.30 & 40.10 & 3.72 & 0.75\\
\quad LTA & 45.35 & 17.27 & 3.74 & 0.39\\[3pt]
Community snapshot &&&&\\
\quad 2006-07-05 (summer) & 34.24 & 11.49 & 3.47 & 0.39\\
\quad 2006-10-25 (fall) & 75.09 & 23.47 & 4.27 & 0.32\\
\quad 2007-01-31 (winter) & 95.29 & 26.99 & 4.51 & 0.30\\[3pt]
Home outdoor & 45.65 & 28.30 & 3.63 & 0.64\\
\hline
\end{tabular*}
\end{table}

\subsection{GIS}

In addition to the monitoring data, spatial prediction at locations
where there are no measurements rely heavily on GIS-based covariates
and so-called ``land-use regression'' techniques [\citet
{Jerrett2005fk}]. In this paper, we considered a limited set of
geographic covariates: (i) log distance to A1, A2 or A3 roadway [\citet
{Teleatlas}], (ii) log Caline3QHCR point predictions averaged over 9\vadjust{\goodbreak}
kilometer buffer [\citet{eckhoff1995addendum}], (iii) distance to
nearest coast [\citet{Teleatlas}], (iv) distance to city hall [\citet
{Teleatlas}], (v) normalized difference vegetation index averaged over
250 meter buffer [\citet{NDVI}], (vi) log elevation, and (vii) percent
impervious surface in 50 meter buffer [\citet{LandCover}].\vadjust{\goodbreak}

\section{Methods}

\subsection{Review of full-rank spatio-temporal model}

The existing spatio-\break temporal model as initially described by Szpiro
[\citet{Szpiro2009}] takes the form
\[
y(s,t) =  \mu(s,t) + \nu(s,t),
\]
where $y(s,t)$ is the log two-week average of pollutant measurements at
location $s$ and time $t$, $\mu(s,t)$ is the mean field and $\nu(s, t)$
is the residual field. The mean field, $\mu$, is defined as a linear
combination of temporal basis functions with spatially varying
coefficients. The spatially varying coefficients are comprised of a
land-use regression component in addition to spatially structured
random fields. These coefficients capture spatial heterogeneity in the
amplitude of the temporal basis functions. As such, the mean field is
written as
\[
\mu(s,t) =\sum_{j=1}^m \bigl\{
\bbm{X}_j\bmm{\alpha}_j+\beta_j(s) +
\psi_j(s)\bigr\}f_j(t),
\]
where the $\bbm{X}_j$ are design matrices containing GIS/land-use
covariates of dimension $n \times(p_j+1)$, where $n$ is the total
number of observed sites and $\bmm{\alpha}_j$ is a vector of regression
land-use regression coefficients of dimension $p_j +1 \times1$. The
$\beta_j(\bbm{s})$ where $\bbm{s} = (s_1, \ldots, s_n)$ are Gaussian
spatial random fields distributed as
\[
\beta_j(\bbm{s}) \sim N\bigl(\bbm{0}, \Sigma_{\beta_j}(\bmm{
\theta}_j)\bigr).
\]
Here, $\Sigma_{\beta_j}(\bmm{\theta}_j)$ is the covariance matrix of
dimension $n\times n$ indexed by the vector of parameters $\bmm{\theta
}_j$. Generally, we assume a spatial exponential decay model with range
$\phi_j$ and partial sill $\tau^2_j$. The $\psi_j(\bbm{s})$ are i.i.d.
random effects distributed as
\[
\psi_j(\bbm{s}) \sim N\bigl(\bbm{0}, \sigma^2_j
\bbm{I}\bigr).
\]
Note $\psi_j(\bbm{s})$ can equivalently be thought of as the nugget for
the $\beta_j(\bbm{s})$-field. The original formulation of this model did
not include a provision for a nugget [\citet{Szpiro2009}], although more
recent work allowed for but did not utilize this parameter [\citet
{Lindstrom2012}]. We later discuss the implications of excluding the
nugget for computation and predictive performance.

\begin{figure}

\includegraphics{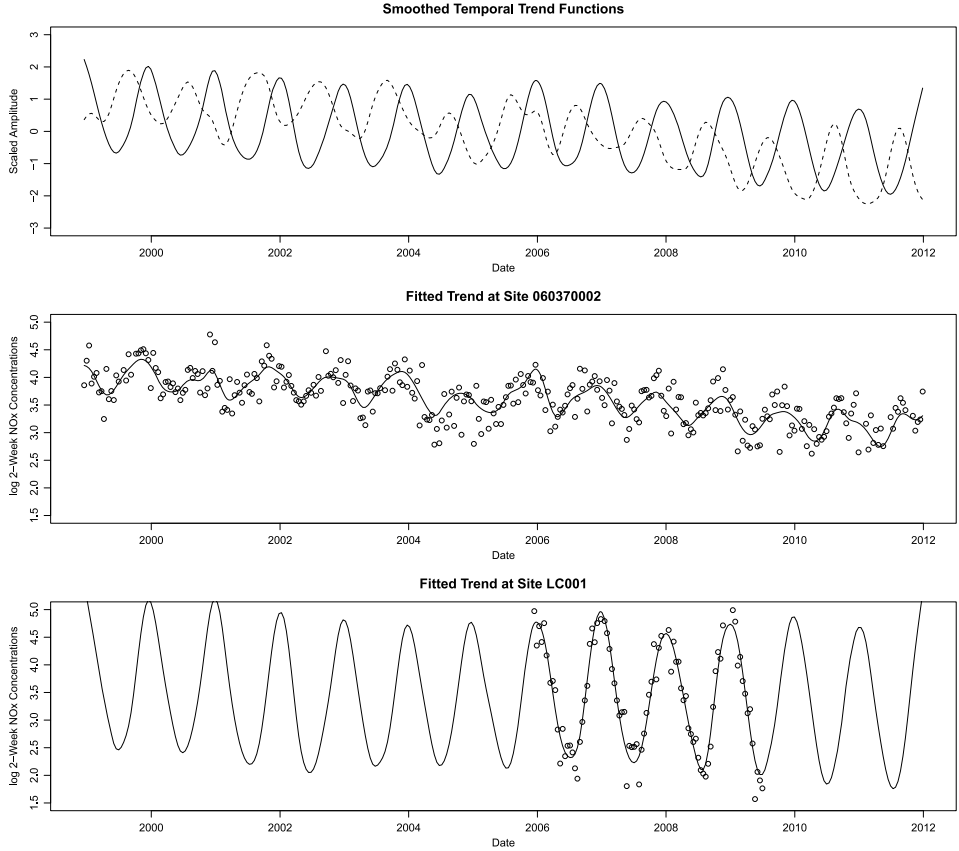}

\caption{(Top) Two temporal basis functions estimated by modified
singular value decomposition from Los Angeles monitoring data; (middle
and bottom) raw log-transformed data and fits to the two temporal basis
functions at sites near (LC001) and far (06037002) from the coastline.}
\label{figtrends}
\end{figure}

The $f_j(t)$ are temporal basis functions with $f_1(t) \equiv1$ for
all $t$ (typically $m$ is small, ${\leq}3$) estimated by modified
singular value decomposition. See \citet
{fuentes2006using,Szpiro2009,Sampson2011} for a more thorough
discussion of trend estimation. Figure~\ref{figtrends} depicts these
smooth temporal basis functions and their fit to the EPA and MESA Air
NO$_x$ monitoring data at two sites.

Last, we specify the model for the residual field, $\nu(s,t)$.
Consistent with \citet{Lindstrom2012,Szpiro2009,Sampson2011}, we
assume that the mean model accounts for the mean structure and all
temporal correlation. Thus, the spatio-temporal residuals are assumed
to have zero mean and to be independent in time, so that
\[
\nu(\bbm{s},t) \sim N\bigl(\bbm{0}, \Sigma_\nu^t(\bmm{
\theta}_\nu)\bigr),
\]
where $\Sigma_\nu^t(\bmm{\theta}_\nu)$ is a covariance matrix of
dimension $n_t \times n_t$ and $n_t$ is the number of sites observed at
time $t$ with $\sum_t n_t = N$, the total number of observations. Once
again, we assume that the $\nu$ field follows a spatial exponential
decay model with range $\phi$, partial sill $\tau^2$ and (possibly)
nugget $\sigma^2$.

A concise representation of this model is given as
%
\begin{equation}\label{eqmod1}
\bbm{Y} = \bbm{F}\bbm{X}\bmm{\alpha} + \bbm{F}\bbm{B} + \bbm{F}\bbm{P}+
\bbm{V},
\end{equation}
where $\bbm{Y}$ is an $N\times1$ vector of stacked responses $y(s, t)$
(first varying $s$ then $t$), $\bbm{F} = (f_{st,is'})$ is an $N \times
mn$ matrix that has elements
\[
f_{st, is'} =\cases{f_i(t), &\quad $\mbox{if }s=s'$,
\cr
0, & \quad\mbox{else},}
\]
$\bbm{X}$ is a block\vspace*{1pt} diagonal matrix with diagonal blocks $\{\bbm{X}_j\}
_{j=1}^m$, $\bmm{\alpha}$ is an\break $\sum_{j=1}^m\{ p_j + 1 \} \times1$
stacked vector of the $\bmm{\alpha}_j$, $\bbm{B}$ is an $mn\times1$
vector of the stacked ${\beta}_j$, $\bbm{P}$ is an $mn\times1$ vector
of the stacked nuggets, $\psi_j$, and $\bbm{V}$ is an $N \times1$
vector of the stacked~${\nu}$ (first varying $s$ then $t$). This model
is thus indexed by the land use regression coefficients, $\bmm{\alpha}$,
and the covariance parameters
\begin{eqnarray*}
\bmm{\theta}_B &=& (\bmm{\theta}_1, \ldots, \bmm{
\theta}_m),\qquad \bmm{\theta }_j=\bigl(\phi_j,
\tau^2_j\bigr),\qquad j=1, \ldots, m,
\\
\bmm{\theta}_P &=& \bigl(\sigma^2_1,
\ldots, \sigma^2_m\bigr),
\\
\bmm{\theta}_V &=& \bigl(\phi_\nu, \tau^2_\nu,
\sigma^2\bigr).
\end{eqnarray*}
To simplify notation, we collect the covariance parameters into the
vector $\bmm{\Xi} = (\bmm{\theta}_B, \bmm{\theta}_P, \bmm{\theta}_V)$. In
the remainder of the manuscript, for the sake of brevity we suppress
the dependence of covariance matrices on their respective parameters,
except where an explicit dependence is illustrative.

Model (\ref{eqmod1}) is typically fit using profile maximum likelihood
methods, although full maximum likelihood and restricted maximum
likelihood approaches are also possible [\citet{Lindstrom2012}]. Sampson
used a multi-stage ``pragmatic'' approach to fitting (\ref{eqmod1}) and
generating predictions [\citet{Sampson2011}]. Lindstr\"{o}m adapted the
model to allow for time-varying covariates, although this extension is
not presented here [\citet{Lindstrom2012}]. This model is implemented in
the \texttt{R}-package, \texttt{SpatioTemporal}, available at \url{http://cran.r-project.org/package=SpatioTemporal}.

\subsection{Motivation for reduced-rank spatial smoothing}
\label{secmotivation}
Although the above formulation of the model has been successful for
predicting air pollution concentrations, we note two limitations of
this formulation, particularly with respect to the $\beta$-fields.
First, we note that it is not natural to interpret the $\beta$-fields
as random effects since it is difficult to imagine the data generating
mechanism that might give rise to such fields [\citet{Hodges2011,Hodges2013}]. Second, the range parameters in the $\beta$-fields tend
to be challenging to estimate in practice. Moreover, Zhang showed that
in the case of spatial generalized linear mixed models, this quantity
is not consistently estimable [\citet{zhang2004inconsistent}].\vadjust{\goodbreak}

As such, we consider a spline-based representation of the $\beta$-fields in the mean model. To motivate, we note that the Gaussian
spatial $\beta$-fields, as defined above, can be represented as spatial
splines as follows. Let
\[
\Sigma_{\beta_j} = \tau^2_{j} \bmm{\Omega},
\]
where $\bmm{\Omega}$ is a matrix such that
\[
\bmm{\Omega} = \bigl\{C\bigl(\|\bbm{s}_i - \bbm{s}_j \|\bigr)
\bigr\}_{i,j \in\mathcal{S}}
\]
and $\mathcal{S}$ is the set of observed spatial locations. For the
exponential model, $C(r) = \exp\{-|r|/\phi\}$. It follows that the $\beta
$-fields can be expressed as
\[
\beta_j(\bbm{s}) = \bmm{\Omega}^{1/2}\bmm{
\delta}_j,
\]
where $\bmm{\delta}_j \sim \operatorname{MVN}(\bbm{0}, \tau^2_j \bbm{I})$. The $n$
columns of the matrix $\bmm{\Omega}^{1/2}$ represent $n$ spatial basis
functions indexed by the parameter $\phi_j$. Written as such, the $\beta$-fields can be viewed as random linear combinations of spatial basis
functions. Exploiting the connection between linear mixed models and
penalized splines, we can view the $\beta_j$-fields as penalized
spatial splines with smoothing parameters $\sigma^2/\tau^2_j$ [\citet
{Ruppert2003}]. Having represented the $\beta$-fields as penalized
splines, it is natural to consider penalized reduced-rank splines
instead as a means of improving model performance and computational efficiency.

We note that an analogous argument can be made for the residual field.
However, it is also the case that the $\nu$-field is well understood
within the traditional framework of random effects models. That is, the
$\nu$-field captures extra random spatial variation that arises from
time point to time point. Furthermore, the range parameter in the $\nu$-field tends to be more stably estimated in practice due to the
repeated measurements over time.

In the following sections, we describe reduced-rank representations of
the \mbox{$\beta$-}fields using low-rank kriging and thin plate regression splines.

\subsection{Low-rank kriging}
\label{seclrk}
We follow the approach outlined by \citet{Kammann2003zr} and \citet
{Ruppert2003} for low-rank\vspace*{1pt} kriging (LRK) of the $\beta$-fields.
Specifically, LRK is achieved by replacing $\bmm{\Omega}$ with $\bbm{Z}\tilde{\bmm{\Omega}}{}^{-1}\bbm{Z}^\top$,
where
\[
\bbm{Z} =  \bigl\{C\bigl(\|\bbm{s}_i - \bmm{\kappa}_j \|\bigr)
\bigr\}_{i \in\mathcal{S}, j
\in\mathcal{K}},\qquad \tilde{\bmm{\Omega}} = \bigl\{C\bigl(\|\bmm{
\kappa}_i - \bmm{\kappa}_j \|\bigr)\bigr\}_{ i,j \in\mathcal{K}},
\]
and $\mathcal{K}$ is the set of spatial knot locations, $\bmm{\kappa}$,
of cardinality $K \ll n$. It follows that we can approximate $\beta_j(\bbm{s})$ by
$\bbm{Z} \tilde{\bmm{\Omega}}{}^{-1/2} \bmm{\delta}_j$, where $\bmm{\delta}_j$ is now a $K$-vector distributed as $\operatorname{MVN}(0, \Sigma_{\delta
_j} = \tau^2_j \bbm{I})$. We note that this approach bears strong
resemblance to the predictive processes presented by Banerjee [\citet
{Banerjee2008}]. In fact, Banerjee noted that LRK is a re-projection of
his predictive process. As such, these approaches are computationally
identical despite the fact that the predictive process is derived
formally from a full-rank parent process.

Letting $\bbm{Z}_B =  \{\bbm{Z} \tilde{\bmm{\Omega}}{}^{-1/2} \}
_{j=1}^m$ and $\tilde{\bbm{B}}$ be the stacked vector of $\bmm{\delta
}_j$s, we can express the spatio-temporal model as
%
\begin{equation}\label{eqmod2}
Y = \bbm{F}\bbm{X}\bmm{\alpha} + \bbm{F}\bbm{Z}_B\tilde{\bbm{B}} +
\bbm{F}\bbm{P}+ \bbm{V}.
\end{equation}
Model (\ref{eqmod2}) can be re-expressed as
\[
\bbm{Y} \sim \operatorname{MVN} ( \bbm{F}\bbm{X} \bmm{\alpha}, \tilde{\Sigma} ),
\]
where
\[
\tilde{\Sigma} = \bbm{F}\bbm{Z}_{B}\Sigma_{\tilde{B}}
\bbm{Z}_{B}^\top\bbm{F}^\top+ \bbm{F}
\Sigma_P\bbm{F}^\top+ \Sigma_V ,
\]
$\Sigma_{\tilde{B}}$ is a block-diagonal matrix with diagonal elements
$\{\Sigma_{\delta_j}\}_{j=1}^m$, $\Sigma_P$ is a block-diagonal matrix
with diagonal elements $\{\sigma^2_j \bbm{I}_n\}_{j=1}^m$, and $\Sigma
_V$ is a block-diagonal matrix with diagonal elements $\{\Sigma_\nu^t\}
_{t=1}^T$. The log-likelihood is given by
\[
l(\bmm{\alpha}, \bmm{\Xi} \mid\bbm{Y}) \propto - \log |\tilde{\Sigma} | - (
\bbm{Y}- \bbm{F}\bbm{X} \bmm{\alpha} )^\top\tilde{\Sigma}^{-1}
(\bbm{Y}- \bbm{F}\bbm{X}\bmm{\alpha} ).
\]
Consistent with \citet{Szpiro2009,Lindstrom2012}, we estimate
regression coefficients $\bmm{\alpha}$ using the profile maximum
likelihood. It is easy to show that
\[
\hat{\bmm{\alpha}} = \bigl(\bbm{X}^\top\bbm{F}^\top\tilde{
\Sigma}^{-1} \bbm{F}\bbm{X}\bigr)^{-1} \bbm{X}^\top
\bbm{F}^\top\tilde{\Sigma}^{-1}\bbm{Y},
\]
so that the profile log likelihood is simplified to
%
\begin{equation}\label{eqprofll}
l_p(\bmm{\Xi} \mid \bbm{Y} ) \propto-\log |\tilde{\Sigma} | - (
\bbm{Y}- \bbm{F}\bbm{X}\hat{\bmm{\alpha}} )^\top\tilde{
\Sigma}^{-1} (\bbm{Y}- \bbm{F}\bbm{X} \hat{\bmm{\alpha}} ),
\end{equation}
and the remaining parameters are estimated as those quantities that
maximize (\ref{eqprofll}). We estimate all parameters using the
L-BFGS-B algorithm as implemented in the \texttt{optim} function in
\texttt{stats} package in \texttt{R}. This is an iterative method that
allows for box constraints on all parameters [\citet{Byrd1995}].

Prediction is achieved by assuming a joint distribution between
observed data~$\bbm{Y}$ and unobserved data $\bbm{Y}^*$,
\[
\pmatrix{\bbm{Y}
\cr
\bbm{Y}^*}
  \sim \left( \pmatrix{
\bbm{F}\bbm{X}
\cr
\bbm{F}^*\bbm{X}^* }
 \bmm{\alpha}, \left[
\matrix{ \tilde{\Sigma} & \tilde{\Sigma}_{\cdot *}
\cr
\tilde{\Sigma }_{*\cdot}& \tilde{\Sigma}_{**} }
 \right] \right),
\]
where $\tilde{\Sigma}_{**}$ is the covariance of $\bbm{Y}^*$ and $\tilde
{\Sigma}_{\cdot*}$ is the cross covariance of $\bbm{Y}$ and $\bbm{Y}^*$.
Predictions are based on the conditional expectation $E[\bbm{Y}^* | \bbm{Y}]$ with MLEs plugged in, namely,
\[
\hat{\bbm{Y}}^* = \bbm{F}^*\bbm{X}^* \hat{\bmm{\alpha}} + \hat{\tilde{
\Sigma}}_{*\cdot}\hat{\tilde{\Sigma}}{}^{-1} (\bbm{Y} - \bbm{F}
\bbm{X} \hat{\bmm{\alpha}} )
\]
with conditional prediction variance
\[
V\bigl(\bbm{Y}^* \mid \bbm{Y}, \hat{\Xi}, \hat{\bmm{\alpha}}\bigr) = \hat{\tilde
{\Sigma}}_{**} - \hat{\tilde{\Sigma}}{}^\top_{*\cdot}
\hat{\tilde{\Sigma }}{}^{-1} \hat{\tilde{\Sigma}}_{\cdot*}.
\]
A drawback of LRK is the dependence of the basis functions on the range
parameters $\phi_j,   j=1, \ldots, m$. Kammann and Wand, for purely
spatial data, address this issue by fixing the value of this parameter
at the maximum spatial distance observed in the data [\citet
{Kammann2003zr}]. Although it is attractive to condition on fixed
spatial basis functions, arbitrary selection of these parameters could
lead to worse predictive performance. The range parameters can be
estimated from the data, albeit at the expense of more challenging
numerical optimization and with the caveat that they may not be
consistently estimable [\citet{zhang2004inconsistent}].

An alternative approach which sidesteps these issues and leverages the
spatial spline formulation calls for the use of alternative spline
bases. Thin plate regression splines are a popular alternative, and we
explore their application in the current problem below.

\subsection{Summary of thin plate regression splines}

Thin plate regression\break splines (TPRS) present an alternative to the LRK
approach and mitigate the issue of estimating the range parameter(s)
[\citet{Wood2003uq}]. Although these models are widely used
(implementation is available in the \texttt{R} package \texttt{mgcv},
e.g.), we briefly summarize the approach with the goal of
describing parallels between TPRS and LRK.

Assume that we wish to estimate the function $f$ based on (purely
spatial) observations $\bbm{Y}$ at locations $\bbm{s} = (s_1, s_2)$ such that
\[
Y_i = f(s_i) + \varepsilon_i
\]
by minimizing this penalized objective function
\[
\bigl\| \bbm{Y} - f(\bbm{s})\bigr\| + \lambda\int_{s_1}\! \int
_{s_2} \biggl( \frac
{\partial^2 f}{\partial s_1^2}+ \frac{\partial^2 f}{\partial s_2^2}+
\frac{\partial^2 f}{\partial s_1\, \partial s_2} \biggr)^2 \,d s_1 \,d s_2.
\]
It can be shown that the solution is given by
%
\begin{equation}\label{eqtprsfull}
f(s) = \sum_{i=1}^n \zeta_i
\eta\bigl(\| s - s_i \|\bigr) + \sum_{j=1}^3
\gamma_j \iota_j(s),
\end{equation}
where\vspace*{2pt} the $\iota_j$ are linearly independent polynomials spanning the
space of polynomials in $\mathcal{R}^2$ (of degree less than $2$) and $
\eta(r) = 2^{-3} \pi^{-1} r^{2}\log(r)$. Further, $\bmm{\zeta}$ and
$\bmm{\gamma}$ are fixed unknown coefficients subject to the constraint
$\bbm{T}^\top\bmm{\zeta} = 0$ with $T_{ij} = \iota_j(s_i)$ [\citet
{green1994nonparametric}].

Let $\bbm{E}$ be a matrix so that $E_{ij} = \eta(\| s_i - s_j \|)$. Wood
presents a reduced-rank approximation of this problem, which is the
solution to the unconstrained optimization problem
\[
\operatorname{minimize} \bigl\| \bbm{Y} - \bbm{U}_K\bbm{D}_K
\bbm{W}_K \bmm{\zeta}^* - \bbm{T} \bmm{\gamma} \bigr\| + \lambda\bmm{
\zeta}^{*\top} \bbm{W}_K^\top\bbm{D}_K
\bbm{W}_K\bmm{\zeta}^*,
\]
where\vspace*{1pt} $\bmm{\zeta}^*$ is a $K-3\times1$ vector of fixed unknown
coefficients, $\bbm{U}\bbm{D}\bbm{U}^\top$ is the eigendecomposition of
$\bbm{E}$ so that the $n$ columns of $\bbm{U}$ are equal to the
eigenvectors of $\bbm{E}$ ordered by their associated eigenvalues from
largest to smallest, $\bbm{D}$ is a diagonal matrix of these
eigenvalues, $\bbm{U}_K$ is a matrix of the first $K$ columns of $\bbm{U}$, and $\bbm{D}_K$ is a matrix of the first $K$ rows and columns of
$\bbm{D}$. Last, $\bbm{W}_K$ is a $K\times K-3$ orthogonal column basis
such that $\bbm{T}^\top\bbm{U}_K \bbm{W}_K= \bbm{0}$ (to account for the
constraint) [\citet{Wood2003uq}].

It is easy to see that this unconstrained optimization is equivalent to
fitting the linear mixed model
\[
\bbm{Y} = \bbm{T}\bmm{\gamma} + \bbm{U}_K\bbm{D}_K
\bbm{W}_K\bmm{\zeta}^* + \bmm{\varepsilon},
\]
where $\bmm{\zeta}^* \sim \operatorname{MVN}(\bbm{0}, \sigma^2_\zeta( \bbm{W}_K^\top\bbm{D}_K\bbm{W}_K)^{-1})$, $\bmm{\varepsilon} \sim \operatorname{MVN}(\bbm{0}, \sigma
^2_\varepsilon\bbm{I})$, and $\lambda= \sigma^2_\varepsilon/ \sigma^2_\zeta
$. Equivalently, let $\bmm{\zeta}^* = ( \bbm{W}_K^\top\bbm{D}_K\bbm{W}_K)^{-1/2} \bmm{\delta}^*$, where $\bmm{\delta}^* \sim \operatorname{MVN}(0, \sigma
^2_\zeta\bbm{I})$, then the above equation becomes the following:
\[
\bbm{Y} = \bbm{T}\bmm{\gamma} + \bbm{U}_K \bbm{D}_K
\bbm{W}_K \bigl(\bbm{W}_K^\top
\bbm{D}_K\bbm{W}_K\bigr)^{-1/2}\bmm{\delta}^* +
\bmm{\varepsilon}.
\]

\subsection{Formulation of \texorpdfstring{$\beta$}{beta}-fields as thin plate regression splines}

We consider modeling the $\beta$-fields as TPRS using the relationship
between penalized splines and mixed models [\citet{Ruppert2003}].
Following the above formulation, we can approximate $\beta_j(\bbm{s})$
in (\ref{eqmod1}) as $\bbm{T}\bmm{\gamma}_j + \bbm{Z}^* \bmm{\delta}_j^*$,
where $\bbm{T}$ contains the spatial coordinates of the monitoring
locations, $\bmm{\gamma}_j$ is a $2 \times1$ vector of fixed unknown
coefficients, $\bbm{Z}^* = \bbm{U}_{K}\bbm{D}_{K}\bbm{W}_{K}(\bbm{W}_{K}^\top\bbm{D}_{K}\bbm{W}_{K})^{-1/2}$, and $\bmm{\delta}^*_{j}$ is
a $K-3\times1$ vector distributed as $\operatorname{MVN}(\bbm{0}, \tau^2_j \bbm{I})$.

We can succinctly incorporate this approximation into our modeling
framework as follows. First, augment the design matrices $\bbm{X}_j$ by
appending the matrix $\bbm{T}$ so that $\bbm{X}^*_j = (\bbm{X}_j \ \bbm{T})$
for $j=1, \ldots, m$ (if $\bbm{X}_j$ already contains the spatial
coordinates as predictors, then this step is unnecessary).
Additionally, append the vector $\bmm{\gamma}_j$ to the $\bmm{\alpha}_j$
so that $\bmm{\alpha}^{*\top}_j = (\bmm{\alpha}_j^\top  \  \bmm{\gamma
}^\top_j)$ for $j=1, \ldots, m$. Last, letting $\bbm{Z}^*_B$ be a
block-diagonal matrix with diagonal elements $\{ \bbm{Z}^*\}_{j=1}^m$
and $\bmm{\alpha}^*$ and $\tilde{\bbm{B}}^*$ be the stacked vectors of
$\bmm{\alpha}_j^*$ and $\bmm{\delta}_j^*$ for $j=1, \ldots, m$,
respectively, we formulate the TPRS version of the spatio-temporal
model as a linear mixed model, as follows:
%
\begin{equation} \label{eqtprsmm}
\bbm{Y} = \bbm{F}\bbm{X}^* \bmm{\alpha}^* + \bbm{F} \bbm{Z}^*_B
\tilde{\bbm{B}}^* + \bbm{F} \bbm{P} + \bbm{V}.
\end{equation}

We note the similarities between equations (\ref{eqmod2}) and (\ref
{eqtprsmm}). In fact, Nychka showed that thin plate splines are
equivalent to kriging using a generalized covariance function [\citet
{nychkaspatial}]. It is clear that the difference between LRK and TPRS
has to do primarily with the choice of basis functions. However, we
also emphasize that the TPRS bases are not dependent on any additional
(e.g., range) parameters. Estimation of model parameters and prediction
follows as described in Section~\ref{seclrk}.

\section{Computational considerations}

Evaluation of (\ref{eqprofll}) directly is computationally intensive,
with the number of computations growing as $\mathcal{O}(N^3)$. However,
the computational burden can be eased considerably by taking advantage
of the block-diagonal nature of the $\Sigma_B$ and $\Sigma_V$. Namely,
Lindstr\"{o}m showed that reformulation of (\ref{eqprofll}) can reduce
the computational burden to $\mathcal{O}(m^3n^3)$ [\citet
{Lindstrom2012}]. Typically, low-rank models boast a computational
advantage over their full-rank counterparts. Yet, reducing the
computational burden in spatio-temporal data is nuanced. In the
following, we discuss how the formulation of the $\beta$-fields using
either LRK or TPRS impacts computation. We illustrate the computational
burden of calculating (\ref{eqprofll}) by considering the determinant
term $|\tilde{\Sigma}|$, employing a similar reformulation to that
employed in \citet{Lindstrom2012} to exploit the block diagonal nature
of $\Sigma_B$ and $\Sigma_V$. Proofs of the following results and the
corresponding reformulation of the full likelihood in (\ref{eqprofll})
are provided in the Online Supplement [\citet{supp}].

By application of known identities, it can be shown that
%
\begin{eqnarray}
\quad| \tilde{\Sigma} | &=& \bigl|\bbm{F} \bbm{Z}_B \Sigma_{\tilde{B}}
\bbm{Z}_B^\top\bbm{F}^\top+ \bbm{F}
\Sigma_P\bbm{F}^\top + \Sigma_V\bigr|
\nonumber
\\
\label{eqdet1}&=&|\Sigma_{\tilde{B}}| |\Sigma_P||\Sigma_V|\bigl|
\Sigma_P^{-1} + \bbm{F}^\top\Sigma_V^{-1}
\bbm{F}\bigr|
\\
\nonumber
&& {}\times \bigl|\Sigma_{\tilde{B}}^{-1} + \bbm{Z}_B^\top
\bbm{F}^\top \bigl(\Sigma_V^{-1} -
\Sigma_V^{-1} \bbm{F}
\bigl(\bbm{F}^\top\Sigma_P^{-1}
\bbm{F} + \Sigma _V^{-1}\bigr)^{-1}
\bbm{F}^\top\Sigma_V^{-1}\bigr) \bbm{F}
\bbm{Z}_B\bigr|.
\end{eqnarray}
For highly unbalanced data like that which\vspace*{1pt} we typically encounter in
MESA Air, (\ref{eqdet1}) is dominated by the calculation of $|\Sigma
_P^{-1} + \bbm{F}^\top\Sigma_V^{-1} \bbm{F}|$. Computation of this
component grows at $\mathcal{O}(m^3n^3)$, the same rate as the
full-rank model.

As mentioned, the full-rank spatio-temporal model originally published
by Szpiro did not include the nugget, $\bbm{P}$, in the $\beta$-fields
[\citet{Szpiro2009}]. When the nugget is not present, the determinant
$|\tilde{\Sigma}|$ reduces to
%
\begin{eqnarray}
|\tilde{\Sigma}| &=& \bigl|\bbm{F} \bbm{Z}_B \Sigma_{\tilde{B}}
\bbm{Z}_B^\top \bbm{F}^\top+
\Sigma_V\bigr|
\nonumber
\\[-8pt]
\label{eqdet2}
\\[-8pt]
\nonumber
&=& |\Sigma_{\tilde{B}}| |\Sigma_V| \bigl|\Sigma_{\tilde{B}}^{-1}
+ \bbm{Z}_B^\top\bbm{F}^\top
\Sigma_V^{-1} \bbm{F} \bbm{Z}_B \bigr|.
\end{eqnarray}
Interestingly,\vspace*{1pt} in (\ref{eqdet2}), computation will generally be
dominated by calculation of $|\tilde{\Sigma}_B^{-1} + \bbm{Z}_B^\top\bbm{F}^\top\Sigma_V^{-1} \bbm{F} \bbm{Z}_B |$, which grows at $\mathcal
{O}(m^3K^3)$. This makes it clear that, when the nugget is not present,
reducing the rank of the $\beta$-fields can\vadjust{\goodbreak} lead to some improvement in
terms of computation. We note that in the case where the data are more
balanced, it is possible that computation of $|\Sigma_V|$ (or,
equivalently, $\Sigma_V^{-1}$), which grows at $\mathcal{O}(\sum_t
n_t^3)$, will dominate computation in both cases.

\begin{figure}[b]

\includegraphics{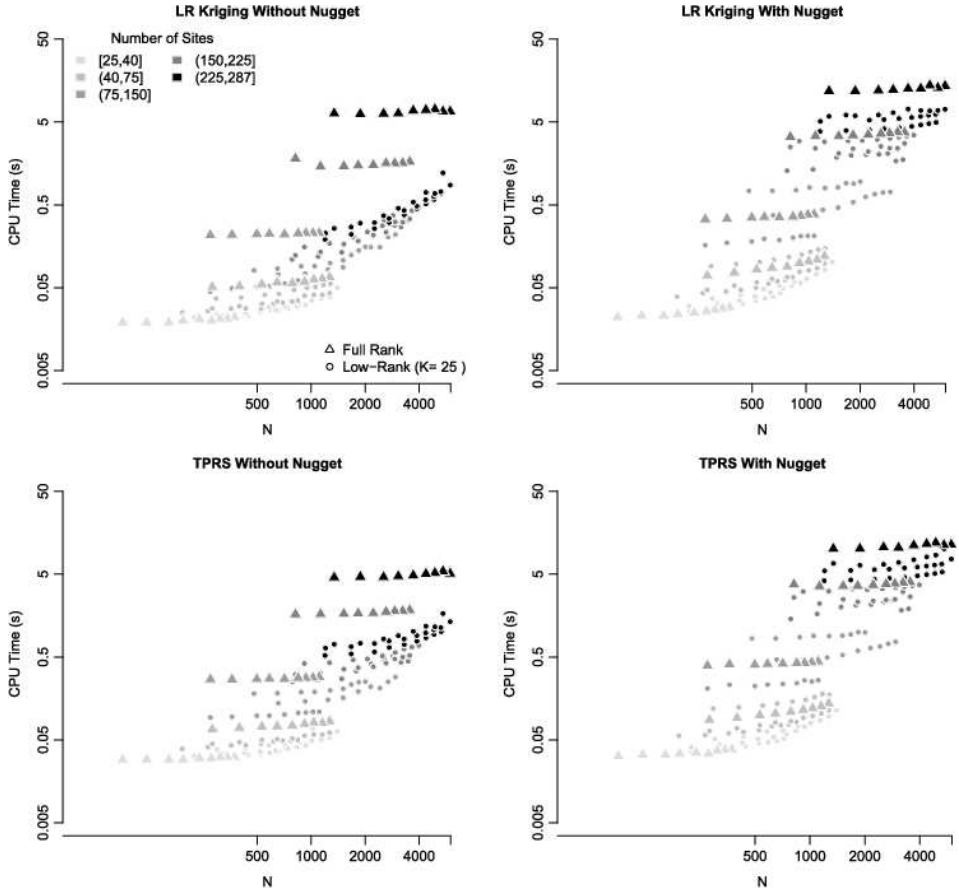}

\caption{CPU time required for a single log-likelihood evaluation of
LRK and TPRS models for the EPA AQS and MESA Air NO$_x$ monitoring data
in Los Angeles, California. Triangles indicate models where the rank of
the spatial smooth is equal to the number of sites and circles indicate
models where the rank of the smooth is equal to 25 in various depleted
MESA Air data sets.}
\label{figcomputation}
\end{figure}

In Figure~\ref{figcomputation}, we plot the CPU time required for
optimized log-likelihood evaluation in full-rank and reduced-rank
models with $K=25$ with and without the nugget present for both LRK and
TPRS. We see that for  LRK and TPRS, as the number of sites
increases, full-rank models take large steps in computation time
required, whereas reduced-rank models grow much more slowly when a
nugget is not present. However, there is very little difference in
computational growth between full- and reduced-rank models as the
number of sites increases when the nugget is present.

\section{Application to \texorpdfstring{NO$_x$}{NOx} monitoring data in Los Angeles}

We apply the proposed reduced-rank spatio-temporal models to NO$_x$
data collected in the Los Angeles area as part of the MESA Air
monitoring campaign and via the EPA regulatory network.

\subsection{Models considered}

We fit a variety of models to the data which vary in three aspects: (1)
the choice of spline basis, (2) the rank of $\beta$-field smooth, and
(3)~the inclusion of the nugget. In all models considered, we employ
two time trends ($m=2$) as depicted in Figure~\ref{figtrends}.
Likewise, the residual $\nu$-field is always specified as exponentially
distributed with a nugget. And, last, all of the GIS covariates are
present in each of the $\bbm{X}_j$ matrices.

\subsubsection{Choice of spline basis}

We have outlined two possible classes of spline bases, exponential
(used in LRK) and thin plate splines. As previously indicated, the use
of exponential basis functions requires handling of the range
parameters in each of the $\beta$-fields by either fixing its value at
some \emph{ad hoc} data-derived value or through full optimization. To
investigate the trade-off between optimization of an additional range
parameter and fixing this parameter at an arbitrary conservative value,
we assume the range parameters in the $\beta$-fields are both fixed,
and in separate models that they are estimated. To assess the
sensitivity to the fixed value, we set the range parameters in all
fields equal to the maximum, one half, one quarter and one eighth of
the observed maximum spatial range in the data (80.7~km). Additionally,
to assess the sensitivity of model performance to the choice of spline
basis, we fit TPRS smooths to the $\beta$-fields.

\subsubsection{Rank of smooth}

As a general rule of thumb, Ruppert, Wand and Carroll suggest that the
number of knots, $K$, be chosen as $\max(20, \min\{150, \break  n/4\})$ [\citet
{Ruppert2003}]. In the case of our MESA Air and EPA data, this would
result in $K=71$. Although this rule of thumb is convenient, it is
unclear how the number of knots in the spatial component of the mean
model will influence spatio-temporal prediction. For our purposes, we
explore a variety of different ranks on spatio-temporal prediction,
$K=287, 100, 50$ and $25$. We note that the models with $K=287$
correspond to full-rank models.

\begin{figure}

\includegraphics{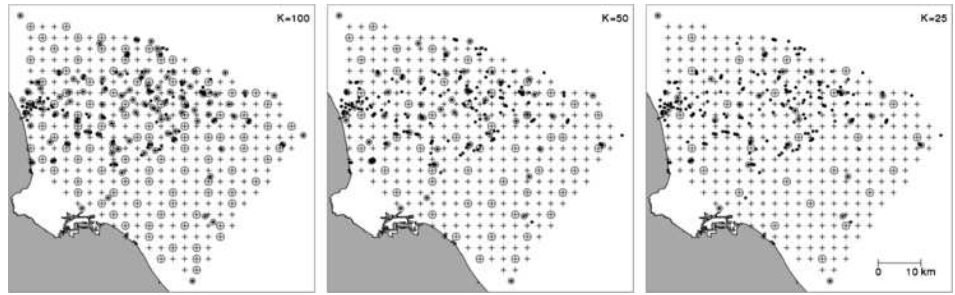}

\caption{Map of knot locations chosen by efficient space-filling of
monitoring locations (small open circle) and of regular grid locations
(large open circles). Small black dots represent participant locations
and crosses represent grid locations. ``Home outdoor'' monitors have
been jittered for participant confidentiality.}
\label{figknots}
\vspace*{-6pt}
\end{figure}

Knot location can also play an important role in LRK. Kammann and Wand
choose knot locations using efficient space-filling algorithms [as
implemented by the \texttt{cover.design()} function in the \texttt{R}
package \texttt{fields}] [\citet{Kammann2003zr}]. In our primary
investigations, we choose knot locations using space-filling of
monitoring sites within the study area (see Figure~\ref{figknots}).
Although space-filling of observed locations is a convenient approach
to choosing the knot locations in our analysis, it is natural to
consider knots chosen at alternative locations. For example, an
attractive option could be to specify knot locations on a regular grid
over the study area. To investigate, in addition to the primary
analysis, we also fit models where knot locations are chosen using
space-filling of a regular grid of the convex hull of the study region,
where each grid cell is approximately 2.5 kilometers on each side (see
Figure~\ref{figknots}).

\subsubsection{Nugget effect}

Given the analytical findings suggesting that\break reduced-rank modeling
leads to a computational advantage in the case when the nugget is not
present, we fit models both with and without the nugget. However, we
note that while analytically feasible, models which exclude the nugget
from the $\beta$-fields are less conceptually defensible. Namely,
exclusion of the nugget from the $\beta$-fields makes it difficult for
the model to capture fine-scale variability in the mean process.
Moreover, preliminary investigations showed that very low-rank smooths
in models without a nugget in the $\beta$-fields were unstable. As
such, we present a limited set of results for reduced-rank models where
the nugget is not present.

\subsection{Model validation}

We employ cross-validation to assess model predictive performance. Our
primary interest is in prediction of long-term averages of NO$_x$
concentrations. Unfortunately, in this data set there are only 26 AQS
and/or MESA ``fixed sites'' that provide adequately long time-series
for long-term average validation. These sites tend to be more
homogeneous in their geographic covariate distribution and have larger
spatial spread when compared to MESA participant locations, which could
potentially limit our ability to adequately assess predictive performance.

As such, in addition to cross-validation of AQS and MESA ``fixed
sites,'' we also consider cross-validation of MESA ``community
snapshot'' and ``home outdoor'' sites. We apply tenfold cross-validation
to each type of monitor. In each of these three scenarios, all
remaining data are used to estimate model parameters and to predict at
left-out locations.

Due to the varying nature of sampling at sites in each of the three
scenarios, Lindstr\"{o}m suggests calculating RMSE and $R^2$ slightly
differently in each case [\citet{Lindstrom2012}]. At ``fixed''/AQS
sites, we calculate RMSE and $R^2$ metrics on both the 2-week and
long-term average scales. Long-term averages at left-out sites are
computed only over times where data are observed, so that
\[
c(s) = \sum_{ \tau \dvtx  \exists y(s, \tau) } \frac{\exp\{y(s, \tau)\}}{|
\{ t \dvtx  \exists y(s, t) \} |}.
\]
The cross-validated $R^2$ on the long-term average scale is given by
[\citet{Szpiro2011fk}]
%
\begin{equation}\label{eqr2}
R^2 = \min \biggl\{0, 1-\frac{\operatorname{RMSE}(\hat{c}(s))^2}{\Var(c(s))} \biggr\}.
\end{equation}

For the second scenario, we perform cross-validation of the ``community
snapshot'' locations. We cross-validate all three sampling
periods/seasons simultaneously and calculate cross-validated RMSE and
$R^2$ by season. Doing so allows us to assess the spatial predictive
ability of the model across multiple seasons. Likewise, as each of the
``community snapshot'' locations were sampled during the same two-week
periods, we can view the resulting metrics as representative of the
pure spatial predictive capacity of the model.

Last, we also consider cross-validation of ``home outdoor'' sites. As
the ``home outdoor'' sites are repeatedly sampled over time and
typically at different time points, much of the $R^2$ is likely to
reflect a temporal signal, which is strong in these data. As such, in
addition to the raw cross-validated $R^2$, we also consider a
de-trended version of the $R^2$ where the variance, $\Var(c(s))$, in
(\ref{eqr2}) is replaced by the variance of observations after
removing the predictions from a reference model that accounts for
(some) temporal variability. Here, we use a reference model based on
the spatial average of measurements at AQS/``fixed sites'' at each time
point. Thus, the de-trended $R^2$ represents the improvement in
performance of our models compared to central site predictions commonly
used in air pollution epidemiology studies [\citet{Pope1995uq}].

\subsection{Comparison with other reduced-rank spatio-temporal models}

Al-\break though the current model was developed specifically to address the
complexities arising in the context of MESA Air, a number of other
methods for reduced-rank spatial and spatio-temporal modeling have been
published, including fixed-rank \mbox{filtering}, Gaussian Markov random field
approximations, covariance tapering, predictive processes and
generalized additive models. Unfortunately, fixed-rank filtering is not
available in an off-the-shelf package, and implementing this model for
these data is a project unto itself. We further note that the
application of Gaussian Markov random field approximations and
covariance tapering in this setting is nuanced and may not result in
any computational savings for these data. See the Online Supplement [\citet{supp}] for
further discussion of the application of these two approaches in the
current modeling framework.

As mentioned previously, there appears to be an explicit correspondence
between predictive processes and LRK, as noted in \citet{Banerjee2008}.
As such, formally modeling the $\beta$-fields in (\ref{eqmod1}) as
reduced-rank predictive processes would not provide any additional
insight into this work. That being said, one version of a predictive
process \emph{spatio-temporal} model is implemented in the \texttt
{spBayes} package in \texttt{R}. Namely, the function \texttt{spDynLM}
fits the following model:
\begin{eqnarray*}
y(s, t) &=& \bbm{X}_t(\bbm{s})\beta_t + u_t(
\bbm{s}) + \varepsilon_t(\bbm{s}),\qquad t=1,2,\ldots, T,
\\
\varepsilon_t(\bbm{s}) &\sim& N\bigl(0,\tau_{t}^2
\bigr),
\\
\beta_t &=& \beta_{t-1} + \eta_t,
\qquad\eta_t \sim N(0,\Sigma_{\eta}),
\\
\beta_0 &\sim& N(m_0, \Sigma_0),
\\
u_t(\bbm{s}) &=& u_{t-1}(\bbm{s}) + w_t(
\bbm{s}),\qquad w_t(\bbm{s}) \sim \operatorname{GP}\bigl(0, C_t(\cdot,
\theta_t)\bigr),
\\
u_0(s) &=& 0.
\end{eqnarray*}
The spatial process $w_t$, here assumed to be exponential, can be
replaced with a predictive process of reduced rank to reduce
computational burden. This model significantly deviates from our own
and may not perform well in the context of such highly imbalanced data
as that which we analyze here. Nevertheless, we apply it to our data
in an effort to make a fair comparison between published approaches to
reduced-rank spatio-temporal modeling and our method. Specifically, we
fit two models:
\begin{longlist}[2.]
\item[1.] full-rank model ($K=287$) for all $w_t$ fields, and
\item[2.] reduced-rank ($K=50$) for all $w_t$ with knots chosen on a grid.
\end{longlist}
We note that the \texttt{spDynLM} function requires that knots be
chosen on a grid when utilizing the reduced-rank predictive process
machinery. In both cases, we fit the models assuming the following
priors for the $\theta_t, \Sigma_\eta, \tau^2_t$:
\begin{eqnarray*}
1/\phi_t &\sim& \operatorname{Unif}\bigl(1/(0.9\times \mbox{max distance}), 3/(0.05 \times
\mbox{max distance})\bigr),
\\
\sigma^2_t &\sim& \operatorname{InvGamma}(2, 10),
\\
\tau^2_t &\sim& \operatorname{InvGamma}(2, 5),
\\
\Sigma^2_\eta &\sim& \operatorname{InvWish}(2, 0.001\bbm{I}_p).
\end{eqnarray*}
These priors are largely based on the example code available in the
\texttt{spDynLM} documentation, with some small changes to reflect the
data. Model predictions were the median of 500 posterior draws, after a
burn-in period of 1500. We cross-validated these models for ``fixed
sites'' using the same cross-validation groups as before.

Last, for an additional comparison with methods available in
off-the-shelf software, we considered a generalized additive model that
reformulates the mean process $\mu(s, t)$ without resorting to a
dynamic model. Namely, we replaced $\mu(s,t)$ with the following:
\[
\bbm{X}(\bbm{s}) \bmm{\alpha} + \eta_t + g(\bbm{s}) + h(\bbm{s}, t).
\]
Here both $g$ and $h$ are modeled using TPRS. For investigating models
with spatial rank of $K$, we set the degrees of freedom for $g$ equal
to $K$ and the degrees of freedom for $h$ equal to $K\times14$ (e.g.,
when $K=50$, $h$ has 700 df), where 14 is the number of years
represented in the data. Note that both $g$ and $h$ can be viewed as
penalized regression splines with structure similar to what we outline
in the paper. But for $h$, we are now assuming a nonseparable model for
space and time which differs from the tensor product approach used in
our model. Moreover, we do not rely on predefined temporal basis
functions to model time. The $\eta_t$ are i.i.d. Gaussian random
effects that capture nonsmooth temporal variation. Note that the $\nu
$-field remains the same as outlined in the paper. We fit this model
using the \texttt{gamm} function in the \texttt{mgcv} package in \texttt{R}.

\begin{table}
\tabcolsep=0pt
\tablewidth=\textwidth
\caption{Cross-validated RMSE and $R^2$ for ``fixed sites'' when nugget
is present. $R^2$ have been multiplied by~100 for presentation}
\label{tabfixed}
\begin{tabular*}{\tablewidth}{@{\extracolsep{\fill}}lcccccccccc@{}}
\hline
& \multicolumn{5}{c}{\textbf{RMSE}} & \multicolumn{5}{c@{}}{$\bolds{R}^{\mathbf{2}}$}\\[-4pt]
& \multicolumn{5}{c}{\hrulefill} & \multicolumn{5}{c@{}}{\hrulefill}\\
\textbf{Basis}$\bolds{/K}$ & \textbf{287} & \textbf{100} & \textbf{50} & \textbf{25} & \textbf{0} & \textbf{287} & \textbf{100} & \textbf{50} & \textbf{25} & \textbf{0}\\
\hline
2-wk\\
\quad LRK ($\phi=\operatorname{est}$)&15.43&15.52&15.72&16.35&17.82&85&85&85&83&80\\
\quad  LRK ($\phi=\max$)&15.64&15.64&15.83&15.87&17.82&85&85&84&84&80\\
\quad LRK ($\phi=\mathrm{max}/2$)&15.52&15.56&15.24&16.14&17.82&85&85&86&84&80\\
\quad LRK ($\phi=\mathrm{max}/4$)&15.31&15.32&15.33&15.74&17.82&85&85&85&85&80\\
\quad LRK ($\phi=\mathrm{max}/8$)&15.04&15.08&15.13&15.59&17.82&86&86&86&85&80\\
\quad TPRS&16.38&15.29&15.11&16.01&17.82&83&85&86&84&80\\[3pt]
LTA\\
\quad  LRK ($\phi=\operatorname{est}$)&10.43&10.56&11.11&11.41&12.41&68&67&64&62&55\\
\quad LRK ($\phi=\max$)&10.48&10.46&10.53&10.84&12.41&68&68&67&65&55\\
\quad LRK ($\phi=\mathrm{max}/2$)&10.40&10.39&10.08&10.72&12.41&68&68&70&66&55\\
\quad LRK ($\phi=\mathrm{max}/4$)&10.30&10.31&10.33&11.04&12.41&69&69&69&64&55\\
\quad LRK ($\phi=\mathrm{max}/8$)&10.26&10.28&10.36&10.85&12.41&69&69&68&65&55\\
\quad TPRS&10.83&\phantom{0}9.99&\phantom{0}9.88&10.60&12.41&65&71&71&67&55\\
\hline
\end{tabular*}
\end{table}

\section{Results}

\subsection{Performance of proposed reduced-rank models in LA}
Table~\ref{tabfixed} shows the results of the cross-validation at
``fixed sites'' for models when the nugget is present. For LRK models,
the choice of range does not appear to be a strong determinant of the
predictive performance, with fully optimized models performing nearly
as well as those models with the range parameter fixed at various
values. Likewise, TPRS models exhibit highly competitive predictive
performance with a slight edge over LRK models at lower ranks for
long-term averages. Cross-validated $R^2$ values stay relatively
consistent across ranks until $K=25$, at which point both 2-week and
long-term average predictive scores drop off. In all cases, models with
some spatial smoothing ($K>0$) perform better than models without any
smoothing ($K=0$).

\begin{table}
\tabcolsep=0pt
\tablewidth=\textwidth
\caption{Cross-validated RMSE and $R^2$ for ``community snapshot''
locations when nugget is present. $R^2$~have been multiplied by 100 for
presentation}
\label{tabcomco}
\begin{tabular*}{\tablewidth}{@{\extracolsep{\fill}}lcccccccccc@{}}
\hline
& \multicolumn{5}{c}{\textbf{RMSE}} & \multicolumn{5}{c@{}}{$\bolds{R}^{\mathbf{2}}$}\\[-4pt]
 & \multicolumn{5}{c}{\hrulefill} & \multicolumn{5}{l@{}}{\hrulefill}\\
\textbf{Basis}$\bolds{/K}$ & \textbf{287} & \textbf{100} & \textbf{50} & \textbf{25} & \textbf{0}& \textbf{287} & \textbf{100} & \textbf{50} & \textbf{25} &\textbf{0}\\
\hline
Summer\\
\quad LRK ($\phi=\est$)&\phantom{0}6.74&\phantom{0}6.71&\phantom{0}6.73&\phantom{0}6.98&\phantom{0}6.97&66&66&66&63&63\\
\quad LRK ($\phi=\max$)&\phantom{0}6.68&\phantom{0}6.67&\phantom{0}7.03&\phantom{0}6.70&\phantom{0}6.97&66&66&63&66&63\\
\quad LRK ($\phi=\mathrm{max}/2$)&\phantom{0}6.69&\phantom{0}6.68&\phantom{0}6.66&\phantom{0}6.96&\phantom{0}6.97&66&66&66&63&63\\
\quad LRK ($\phi=\mathrm{max}/4$)&\phantom{0}6.71&\phantom{0}6.72&\phantom{0}6.66&\phantom{0}6.76&\phantom{0}6.97&66&66&66&65&63\\
\quad LRK ($\phi=\mathrm{max}/8$)&\phantom{0}6.76&\phantom{0}6.74&\phantom{0}6.83&\phantom{0}6.82&\phantom{0}6.97&65&66&65&65&63\\
\quad TPRS&\phantom{0}6.62&\phantom{0}6.71&\phantom{0}6.70&\phantom{0}6.72&\phantom{0}6.97&67&66&66&66&63\\[3pt]
Fall\\
\quad  LRK ($\phi=\est$)&11.64&11.61&11.99&11.55&11.78&75&76&74&76&75\\
\quad LRK ($\phi=\max$)&11.66&11.59&11.75&11.83&11.78&75&76&75&75&75\\
\quad LRK ($\phi=\mathrm{max}/2$)&11.66&11.64&11.61&11.97&11.78&75&75&76&74&75\\
\quad LRK ($\phi=\mathrm{max}/4$)&11.65&11.63&11.53&11.35&11.78&75&75&76&77&75\\
\quad LRK ($\phi=\mathrm{max}/8$)&11.66&11.89&11.95&11.86&11.78&75&74&74&74&75\\
\quad TPRS&11.92&12.10&12.14&12.11&11.78&74&73&73&73&75\\[3pt]
Winter\\
\quad LRK ($\phi=\est$)&13.01&12.92&12.98&12.99&15.32&77&77&77&77&68\\
\quad LRK ($\phi=\max$)&13.04&12.94&13.11&14.00&15.32&77&77&76&73&68\\
\quad LRK ($\phi=\mathrm{max}/2$)&13.03&12.99&12.59&13.63&15.32&77&77&78&75&68\\
\quad LRK ($\phi=\mathrm{max}/4$)&13.02&12.98&12.63&13.8\phantom{0}&15.32&77&77&78&74&68\\
\quad LRK ($\phi=\mathrm{max}/8$)&13.05&13.51&12.65&13.95&15.32&77&75&78&73&68\\
\quad TPRS&13.27&13.04&13.08&14.19&15.32&76&77&77&72&68\\
\hline
\end{tabular*}
\end{table}

Table~\ref{tabcomco} show the results of cross-validation at
``community snapshot'' sites. We typically see the best performance in
the Winter as compared with the Fall and Spring seasons. Once again,
there appears to be little difference in model performance as the
choice of range parameters varies. TPRS models continue to compete
strongly with LRK models. The rank of the $\beta$-field smooth does not
tend to influence performance heavily, although again spatial smoothing
at any rank does tend to improve predictive performance.

\begin{table}
\tabcolsep=0pt
\tablewidth=\textwidth
\caption{Cross-validated RMSE and $R^2$ for ``home outdoor'' locations
when nugget is present. $R^2$ have been multiplied by 100 for presentation}
\label{tabhome}
\begin{tabular*}{\tablewidth}{@{\extracolsep{\fill}}lcccccccccc@{}}
\hline
& \multicolumn{5}{c}{\textbf{RMSE}} & \multicolumn{5}{c@{}}{$\bolds{R}^{\mathbf{2}}$}\\[-4pt]
&\multicolumn{5}{c}{\hrulefill} &   \multicolumn{5}{c@{}}{\hrulefill}\\
\textbf{Basis}$\bolds{/K}$ & \textbf{287} & \textbf{100} & \textbf{50} & \textbf{25} & \textbf{0}& \textbf{287} & \textbf{100} & \textbf{50} & \textbf{25} &\textbf{0} \\
\hline
Raw\\
\quad LRK ($\phi=\est$)&5.51&5.52&5.88&5.89&7.03&93&93&92&92&88\\
\quad LRK ($\phi=\max$)&5.54&5.54&5.71&6.41&7.03&93&93&92&90&88\\
\quad LRK ($\phi=\mathrm{max}/2$)&5.53&5.54&5.28&6.01&7.03&93&93&93&92&88\\
\quad LRK ($\phi=\mathrm{max}/4$)&5.53&5.58&5.52&6.14&7.03&93&93&93&91&88\\
\quad LRK ($\phi=\mathrm{max}/8$)&5.52&5.49&5.48&6.26&7.03&93&93&93&91&88\\
\quad TPRS&5.52&5.50&5.53&6.00&7.03&93&93&93&92&88\\[3pt]
Detrended\\
\quad LRK ($\phi=\est$)&&&&&&84&84&82&82&75\\
\quad LRK ($\phi=\max$)&&&&&&84&84&83&79&75\\
\quad LRK ($\phi=\mathrm{max}/2$)&&&&&&84&84&86&81&75\\
\quad LRK ($\phi=\mathrm{max}/4$)&&&&&&84&84&84&81&75\\
\quad LRK ($\phi=\mathrm{max}/8$)&&&&&&84&84&85&80&75\\
\quad TPRS&&&&&&84&84&84&81&75\\
\hline
\end{tabular*}
\vspace*{-6pt}
\end{table}

Table~\ref{tabhome} shows the results of the cross-validation study at
``home outdoor'' locations. Here, the choice of range parameter model
appears to have even less of important role locations than it did at
``fixed sites.'' Namely, cross-validated RMSE increases only slightly,
resulting in a minimal decrease in $R^2$, as the rank decreases in the
raw home predictions. Detrended $R^2$ did show some decay as the rank
decreased, but still remained relatively high. TPRS models performed as
well as LRK models across ranks. Again, models with some spatial
smoothing outperformed those models with no smoothing.

\begin{figure}

\includegraphics{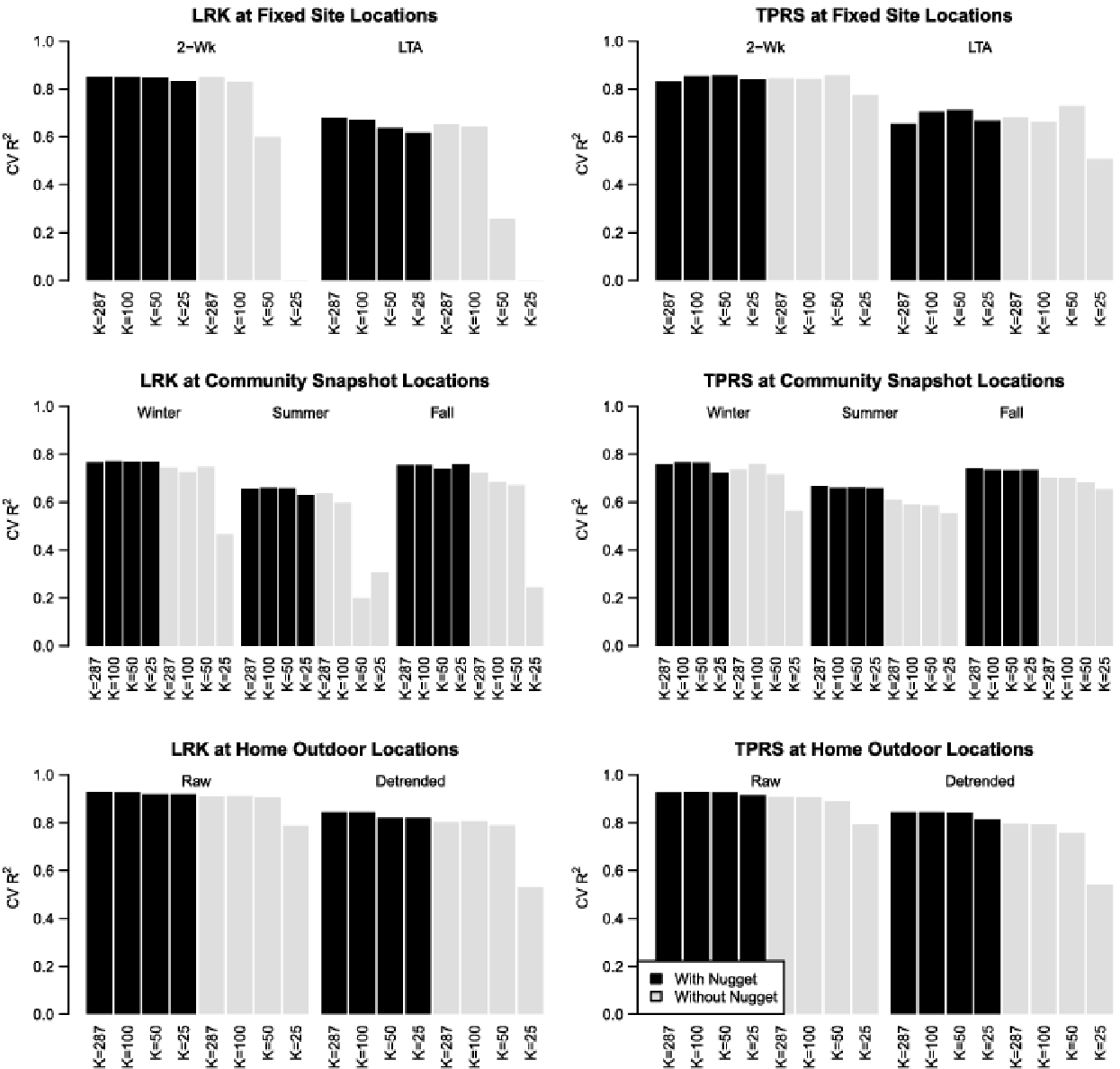}

\caption{Comparison of cross-validated $R^2$ at ``fixed site,''
``community snapshot,'' and ``home outdoor'' locations using low-rank
kriging and TPRS.}
\label{figbarplot}
\end{figure}

Figure~\ref{figbarplot} compares the cross-validated $R^2$ for a set
of models of rank $K=287, 100, 50$ and $25$ with and without the nugget
present in the $\beta$-fields at AQS/``fixed sites,'' ``community
snapshot'' and ``home outdoor'' locations. Note, for LRK results, the
range parameter has been estimated from the data. The figure suggests
that while full rank models ($K=287$) are comparable across these two
specifications, predictive performance of models without the nugget in
the $\beta$-fields tend to drop off rapidly as the rank of the smooth
decreases, particularly in the case of LRK, where in select cases the
$R^2$ decreases to zero when $K=25$. TPRS models tend to be more
robust, although the decrease in $R^2$ in TPRS models without a nugget
tends to be greater than in TPRS models with a nugget.

\begin{figure}

\includegraphics{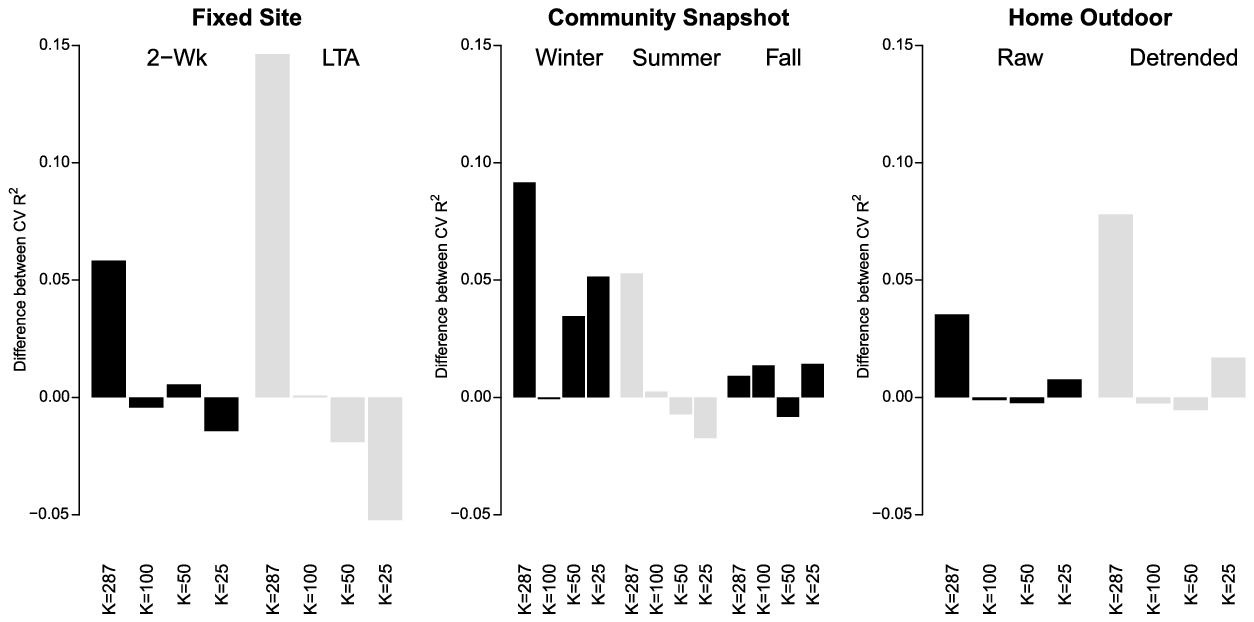}

\caption{Differences between cross-validated $R^2$ in LRK models with
knots chosen at monitoring locations and on a regular grid by rank.
Models assume that the nugget, $\bbm{P}$, is present in all $\beta$-fields and the range parameters are estimated by maximum likelihood.}
\label{figsensknots}
\end{figure}
\begin{figure}

\includegraphics{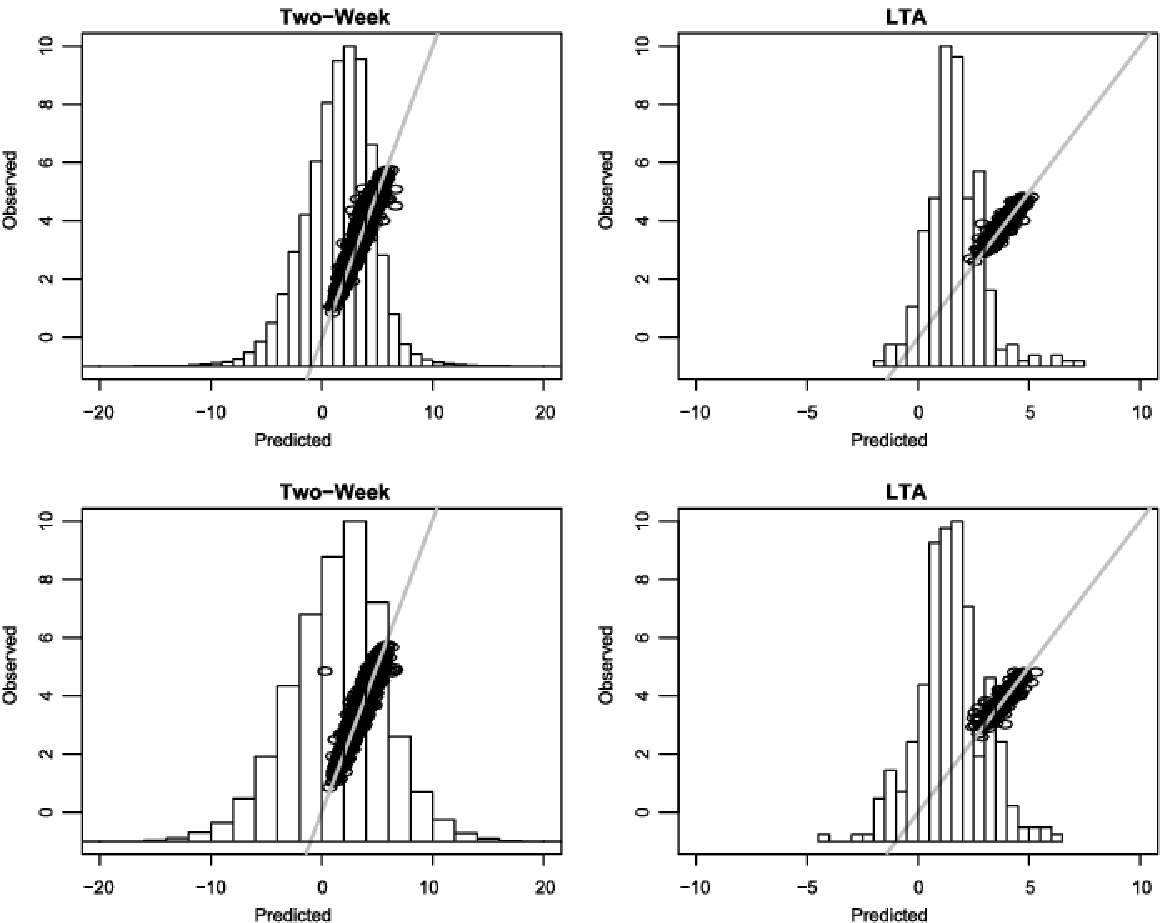}

\caption{(Top row) \emph{In-sample fits} for full-rank models fit in
\texttt{spBayes}. (Bottom row) \emph{In-sample fits} for reduced-rank
models ($K=50$) fit in \texttt{spBayes}. Histograms represent the
distribution of posterior predictions at time points/locations without
observed data.}
\label{figspbayesfit}
\end{figure}

Figure~\ref{figsensknots} compares the results of fitting full and
LRK models to the MESA Air data when the knots were chosen using
space-filling of either monitoring locations or a regularly spaced grid
of locations. Generally speaking, models where knots were chosen at
monitoring sites performed better than those where knots were chosen at
grid locations.

\subsection{Performance of other reduced-rank spatio-temporal modeling methods}

We found that the \texttt{spDynLM} implementation did not work well for
our data, possibly due to the large imbalance across space and time. In
both models ($K=50, 287)$, the time-varying range parameter was not
well identified and varied significantly, thus resulting in poor
characterization of the rate of spatial decay. The temporal sparsity of
the data may also have contributed to the poor performance due to the
dynamic nature of the model's temporal trend. While the in-sample fits
for these models are quite good, the out-of-sample predictions are
highly variable, resulting in cross-validated $R^2$ equal to zero for
all scenarios considered. In Figure~\ref{figspbayesfit} we show the
scatter plots of observed and predicted values in fitted models. The
histograms in this same figure represent the distribution of
predictions at unobserved times and locations. We note that these are
on the log-scale, so that when exponentiated to the native scale, many
predicted values at unobserved times/locations are extremely large.

\begin{figure}

\includegraphics{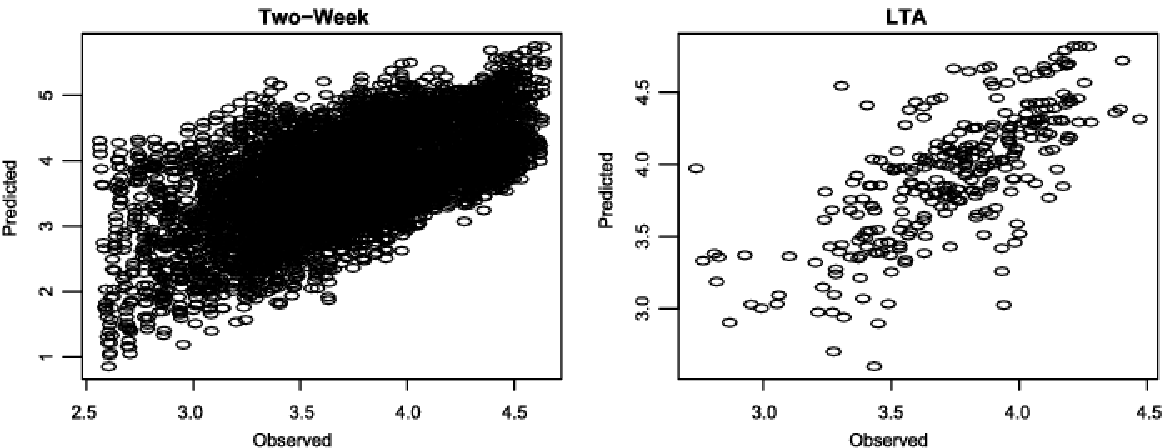}

\caption{\emph{In-sample fits} for reduced-rank models fit in \texttt{mgcv} with $K=50$ on the two-week scale (left) and long-term average
scale (right).}
\label{figgammfit}
\end{figure}

The results of our \texttt{gamm} implementation were only marginally
better. While the in-sample fits of this approach were more promising
(see Figure~\ref{figgammfit}), the predictions were nowhere near the
caliber of those achieved using our model. Inspection of the residuals
suggests that there remains significant temporal correlation that is
unaccounted for by the mean model. We found that the cross-validated
$R^2$ was equal to 0 on both the two-week and long-term average scale
using this approach. This low $R^2$ was driven by the presence of
outlying predictions for a handful of sites in two different
cross-validation groups. Additionally, the model failed to converge for
a single cross-validation group.

\section{Discussion}

This paper focuses on presentation of LRK and TPRS representations of
the mean process in the spatio-temporal model proposed by \citet
{Szpiro2009}, \citet{Sampson2011}, and \citet{Lindstrom2012}. Our
approach allows for a reduced-rank representation of the $\beta$-fields
in the mean process of the original model, which tends to be the most
time-consuming piece to evaluation in likelihood optimization. In
certain cases, we have shown that such reduced-rank representations of
the $\beta$-fields can lead to a computational advantage over the full
rank specification. Namely, when the nugget of the $\beta$-field is not
present, we have shown that our low-rank approach leads to slower
growth in the CPU time required for likelihood evaluation.

The formulation of the $\beta$-fields in the mean process of the model
as spatial splines is attractive for a number of other reasons. For
example, oftentimes predictions of air pollution concentrations are
used as inputs into health models to estimate health effects.
Typically, the predictions are based on spatially misaligned data and
ignoring this fact can lead to biased results and overly optimistic
standard errors [\citet{Szpiro2011fk}]. The expression of the $\beta
$-fields as splines places GIS covariates and spatial smoothing on more
equal footing. Namely, in this form we can think of the GIS covariates
and the spatial basis functions as unpenalized and penalized spatial
covariates, respectively. This interpretation leads to a more coherent
approach to measurement error correction for spatially misaligned data
[\citet{SzpiroPaciorek2012}]. It is also important to note that the
computational advantage gained in log-likelihood evaluation extends
analogously to prediction, thus reducing computation time needed to
predict at potentially many new locations.

For LRK models, we explored the choice of range parameters of
prediction, ranging from the case where the range was fully estimated
from the data to the case where it was fixed at an arbitrary
conservative value indicated by the data. In the scenario when the
range parameter is fixed, we showed that the original specification of
the full-rank model can also be interpreted as a standard penalized
spatial spline.

Likewise, we discussed the parallels between kriging and TPRS. We
emphasize that a limitation of the kriging basis functions is the
reliance on the range parameter and that TPRS is not subject to the
same limitation. That being said, we note that there is an equivalence
between thin plate splines and kriging using a Matern-covariance with
infinite range [\citet{wahba1981spline,nychkaspatial,kimeldorf1970correspondence}]. As such, one might view
the use of TPRS as making an implicit assumption about the range
parameter. The fact that TPRS and LRK were competitive in our results
indicates that TPRS is a valid and attractive option for spatial
smoothing in these models. To further this argument, we performed
additional analyses (results included in the Online Supplement [\citet{supp}])
comparing out-of-sample prediction variances and AIC as a means of
model selection. These analyses indicated that TPRS models tended to
result in more stable prediction variances across rank specification
when compared to LRK models. However, there was little notable
difference between AIC values in LRK and TPRS models. Rather, AIC
values indicated full-rank LRK models were preferable to reduced-rank
ones in all cases. TPRS models with $K=100$ had the lowest AIC.

Our approach to model assessment relies on cross-validation. As we are
primarily interested in prediction of long-term averages, the
cross-validation approach outlined isolates the spatial predictive
capacity of the models. We applied our approach to ambient MESA Air and
EPA NO$_x$ data collected in the Los Angeles area as well as
traditional road covariates and Caline point predictions models. We
found that generally speaking, the choice of the range parameter in the
LRK exponential spatial basis functions had little impact on the model
performance. In fact, reducing the rank of the model tended to also
have little impact in most cross-validation scenarios for ranks of
moderate size ($K=50, 100$). We note that the recommendation of
Ruppert, Wand and Carroll ($K=71$ for the MESA Air data) falls squarely
in this range [\citet{Ruppert2003}]. However, we found that reduction of
the rank of the $\beta$-fields below $K=50$ tended to noticeably impact
model predictions. This impact was further exacerbated by exclusion of
the nugget in the $\beta$-fields. This finding is not a surprise, as
exclusion of the nugget in the $\beta$-fields amounts to attributing
all extra variation in the mean beyond what is explained by the GIS
covariates to the spatial $\beta$-fields. Reduction of the rank of the
smooth of these random fields results in a spatial smooth that is
unlikely to be able to capture spatial heterogeneity.

This unfortunate finding is at odds with the goal of reducing the
computational burden of full-rank spatio-temporal likelihood
evaluations. Although the original specification published by Szpiro et
al. did not include a nugget in the $\beta$-field, it is our feeling
that such models are less defensible than those that include a nugget,
since it is unlikely that the GIS covariates in the model account for
all nonsmooth spatial variation.

That being said, the results herein described are based on a single
data setting. Indeed, there almost surely exists other data sets where
inclusion of a nugget in the $\beta$-fields is contraindicated. In
these cases, use of a moderate rank smooth could lead to both a
computational and predictive advantage.

Last, we examined a number of other approaches and specification to
modeling NO$_x$ concentrations in the current data set and found poor
performance for two off-the-shelf packages. Our findings confirm that
the long history of methodological development of the model under study
in the context of modeling air pollution exposures for MESA Air was
indeed well guided and that current off-the-shelf packages are not
ideal for analyzing these data. Future research should, however,
include investigations into the extension of the current model using
covariance tapering of either the $\beta$-fields covariance or even of
the overall covariance matrix~$\tilde{\Sigma}$.

The LA NO$_x$ data application is meant to exemplify the current
methods. However, we note that this model is being applied more broadly
to four separate pollutants in six major cities in the United States as
part of MESA Air [\citet{Keller2014}]. Furthermore, a rigorous approach
to model selection, that varies the number of trends, covariates and
$\beta$-field models, is also being applied to choose the best
performing predictive models. Taking into account cross-validation,
this effort includes the fitting of hundreds of models, representing a
significant investment of time on the part of MESA Air investigators.
To further emphasize the impact of the current methods, we performed a
separate set of analyses replicating a large subset of the
cross-validation scenarios for NO$_x$ data in Los Angeles considered by
MESA Air investigators in their development of exposure models for use
in primary MESA Air health analyses. We found that TPRS models achieved
highly competitive results in roughly half the time, suggesting that
had these methods been available during model development, potentially
hundreds of computer hours could have been saved during the model
development process. As we move toward incorporating the current
methods into the highly optimized \texttt{SpatioTemporal} package, and
further optimize the reduced-rank model fitting procedures, we expect
that the gains in computational time will increase in orders of
magnitude, to roughly 5 times faster. As such, we believe that the
current work will continue to have tangible implications for MESA Air
investigators and their collaborators who continue to use the MESA Air
spatio-temporal model as the basis for exposure assessment in air
pollution cohort studies.

\section*{Acknowledgments}

This document has not been formally reviewed by the EPA. The views
expressed in this document are solely those of the University of
Washington and the EPA does not endorse any products or commercial
services mentioned in this publication.

\begin{supplement}[id=suppA]
\stitle{Supplement  to ``Reduced-rank
spatio-temporal modeling of air pollution concentrations in the
Multi-Ethnic Study of Atherosclerosis and Air Pollution''}
\slink[doi]{10.1214/14-AOAS786SUPP} 
\sdatatype{.pdf}
\sfilename{aoas786\_supp.pdf}
\sdescription{We provide a detailed derivation of the optimized
likelihood, comparisons of the prediction variances, discussion model
selection by AIC for the paper ``Reduced-rank spatio-temporal modeling
of air pollution concentrations in the Multi-Ethnic Study of
Atherosclerosis and Air Pollution'' by Casey Olives, Lianne Sheppard,
Johan Lindstr\"{o}m, Paul D. Sampson, Joel D. Kaufman and Adam A. Szpiro.}
\end{supplement}


%


\printaddresses
\end{document}